\documentclass[aps,pre,superscriptaddress,twocolumn,longbibliography]{revtex4-2}

\usepackage{graphicx,epstopdf,url}
\usepackage[colorlinks=true,urlcolor=blue,citecolor=blue,linkcolor=blue,urlcolor=blue]{hyperref}

\usepackage[active]{srcltx}

\begin{document}
\title{Localized nonlinear excitations of a columnar chain of coronene molecules
}

\author{Alexander V. Savin}
\email[]{asavin@chph.ras.ru}
\affiliation{Semenov Institute of Chemical Physics, Russian Academy of Sciences,
Moscow 119991, Russia}
\affiliation{Plekhanov Russian University of Economics, Moscow 117997, Russia}

\author{Sergey~V.~Dmitriev}
\email[]{dmitriev.sergey.v@gmail.com}
\affiliation{Institute of Molecule and Crystal Physics, Ufa Federal Research Centre of Russian Academy of Sciences, Oktyabrya Ave. 151, 450075 Ufa, Russia}
\affiliation{Institute of Mathematics with Computing Centre, Ufa Federal Research Centre of Russian Academy of Sciences, Ufa 450008, Russia}

\begin{abstract}
The nonlinear dynamics of a one-dimensional molecular crystal in the form of a chain of planar coronene molecules is analyzed. Using molecular dynamics, it is shown that a chain of coronene molecules supports acoustic solitons, rotobreathers, and discrete breathers. An increase in the size of planar molecules in a chain leads to an increase in the number of internal degrees of freedom. This results in an increase in the rate of emission of phonons from spatially localized nonlinear excitations and a decrease in their lifetime. Presented results contribute to the understanding of the effect of the rotational and internal vibrational modes of molecules on the nonlinear dynamics of molecular crystals.
\end{abstract}

\maketitle

\section{Introduction}

Molecular crystals can have a quasi-one-dimensional morphology, for example, fullerene nanowhiskers consisting of fullerene molecules~\cite{Kausar20221908}, a columnar structure of carbon nanotori~\cite{LunFu20211197,LunFu20222293}, B$_{42}$ molecules~\cite{cryst10060510}, $n$-coronene molecules~\cite{HernRojas201613736,Bartolomei201714330,HernRojas20171884,Chen201479}, columnar discotic liquid crystals~\cite{Feng2009421,Chen201217869,De201918799} and many others. Finite-size particles of molecular crystals have rotational degrees of freedom that can give rize to such cointerintuitive effects as negative thermal expansion~\cite{Wang222105,Koocher053601,Dove094105,Ambrumenil024309,Galiakhmetova2100415} and auxeticity (negative Poisson's ratio)~\cite{Dudek025009,Dudek46529,Gatt201515,Goldstein,VasPav}.

Quasi-one-dimensional crystals can support various spatially localized nonlinear excitations, their study is important and is often considered in connection with the transfer of energy, mass and information. If the molecules that make up quasi-one-dimensional crystals, in addition to translational, also have rotational and internal vibrational degrees of freedom, then the variety of localized excitations supported by them increases.

Let us note the most intensively studied spatially localized excitations in nonlinear lattices and crystals.

{\em Compressive acoustic solitons} are typically excited in solids or metamaterials under shock loading~\cite{Nesterenko2127,Zhang5605,Liu295403,Jian115401}. Acoustic solitons propagating at a speed exceeding the speed of longitudinal sound were described in carbon nanotube bundles~\cite{Galiakhmetova104460}, black phosphorene~\cite{Shepelev115519}, graphene and boron nitride~\cite{Shepelev109549}. It is shown that the attenuation of compressive waves in black phosphorene occurs faster than in graphene and boron nitride due to the greater number of degrees of freedom in the translational cell of phosphorene, which provides more channels for energy emission~\cite {Shepelev115519}.

{\em Rotobreathers} are dynamical modes with a single rotating particle while neighboring particles oscillate with the amplitude decreasing exponentially with distance from the rotating particle~\cite{Takeno140,Takeno1922,Aubry1997201,Martinez1999444}. The works~\cite{Smirnov849,Smirnov123121} are devoted to the analysis of the stability of rotobreathers. The effect of rotobreathers on heat capacity~\cite{Takeno140}, thermal conductivity~\cite{Giardin2144,Gendelman2381}, and slow relaxation~\cite{Eleftheriou230} was analyzed within the framework of one-dimensional rotator lattices. Rotobreathers were considered in a damped driven rotator lattice~\cite{BonartR1134} and in the lattices with geometrical nonlinearities~\cite{Kevrekidis066627,Savin034207}.
The method of molecular dynamics~\cite{Bubenchikov2000174} was used to describe the precession of a rotating fullerene inside a fullerite crystal. The work~\cite{Lunfu2022260} shows the effect of C$_{60}$ fullerite crystal deformation on the rotational dynamics and shift of the center of mass of a single C$_{60}$ molecule. In the works~\cite{Dmitriev2050010,Savin22,Savin116627} rotobreathers in the form of carbon nanotubes rotating around their axis in a carbon nanotube bundle were studied. The dynamics of a fullerene molecule rotating in a fullerite crystal was studied in~\cite{Savin125427}.

{\em Discrete breathers} or {\em intrinsic localized modes} are the large-amplitude, spatially localized vibrational modes in defect-free nonlinear lattices~\cite{Flach1998181,Flach20081,Aubry20061}.
Discrete breathers are ubiquitous in nonlinear lattices and are investigated in models described by the discrete nonlinear Schr\"odinger equation~\cite{Alfimov2004127}, in Josephson superconducting junctions~\cite{Ustinov2003716,Trias2000741}, in granular crystals~\cite{Chong413003}, in a mass-spring chain~\cite{Watanabe20181957}, and in magnetic systems~\cite{Kavitha201691,Bostrem214420,Bostrem015208}. Interatomic interactions are non-linear, so different crystals support discrete breathers~\cite{Dmitriev2016446,Korznikova2016277,Murzaev20171003,Dubinko041124}. In real discrete systems, e.g. in crystals, one deals with quasi-breathers that are not exactly periodic single-frequency modes~\cite{Chechin036608}. A discrete breather in the form of a single fullerene molecule oscillating with a large amplitude in a fullerite crystal~\cite{Savin125427} and a single oscillating carbon nanotube in a nanotube bundle~\cite{Savin116627} were studied by the method of molecular dynamics.

Most popular approaches to the study of nonlinear excitations in molecular crystals are the use of molecular dynamics~\cite{LunFu20211197,LunFu20222293} and coarse-grained models~\cite{HernRojas20171884,HernRojas201613736,Pavlov20221669,ma15144871}.

The aim of this study is to analyze the effect of internal vibrational degrees of freedom on the robustness of various spatially localized nonlinear excitations in a quasi-one-dimensional chain of $n$-coronene molecules with $n=2$, 3, and 4~\cite{HernRojas201613736,Bartolomei201714330,HernRojas20171884}. As the index $n$ increases, the size of the molecules and, consequently, the number of internal degrees of freedom also increase.

In Sec.~\ref{Model}, the structure of the $n$-coronene and the molecular dynamics model used in this study are described. The spectrum of small-amplitude vibrations of the $n$-coronene is analyzed in Sec.~\ref{DispCurves}. Sections from \ref{AcousticSolitons} to \ref{DBs} present the results of studying spatially localized nonlinear excitations in the chains of $n$-coronene molecules, namely, acoustic solitons, rotobreathers, and discrete breathers, respectively. Our conclusions are formulated in Sec.~\ref{Conclusions}.
\begin{figure}[tb]
\begin{center}
\includegraphics[angle=0, width=1.0\linewidth]{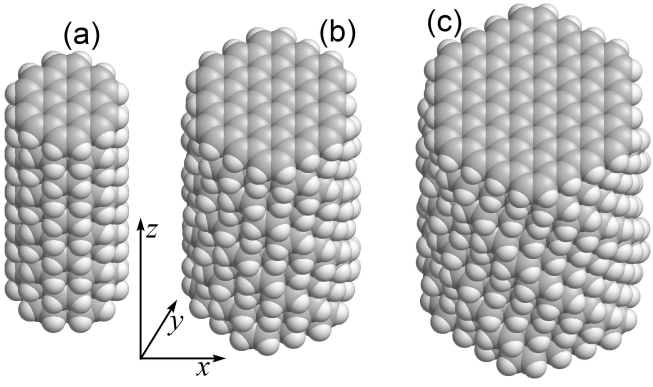}
\end{center}
\caption{\label{fig01}\protect
Vertical chain of 10 $n$-coronene molecules C$_{6n^2}$H$_{6n}$: (a) $n=2$ (coronene C$_{24}$H$_ {12}$); (b) $n=3$ (circumcoronene C$_{54}$H$_{18}$); (c) $n=4$ (dicircumcoronene C$_{96}$H$_{24}$). Carbon atoms (gray) form planar disk molecules, and hydrogen atoms are located at the edges of the disks (shown in light gray). The vertical axis of the chain is parallel to the $z$ axis, the planar molecules are parallel to the $xy$ plane. The positions of neighboring molecules in the chain differ by the shift along the $z$ axis and the relative rotation of the molecules in the $xy$ plane (shift $\Delta z$ and twist $\Delta\phi$ steps of the chain).
}
\end{figure}

\section{Model}
\label{Model}

The $n$-coronene molecule C$_{6n^2}$H$_{6n}$ can be considered as a graphene flake. Therefore, to describe the dynamics of a coronene molecular crystal, one can use the force field previously used for graphene nanoribbons.

To simplify the modeling, valence-bonded CH groups of atoms at the edges of disk molecules will be considered as a single carbon atom of mass $13m_p$, while all other inner carbon atoms have the mass $12m_p$, where $m_p=1.6601\times10^{ -27}$~kg is the proton mass.

\begin{figure}[t]
\includegraphics[angle=0, width=1\linewidth]{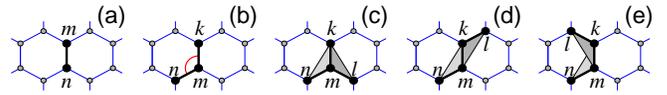}
\caption{(Color online)
Different types of interactions between neighboring atoms belonging to the sets  $\Omega_j$, $j=1$, 2, 3, 4, 5. (a) Valence interactions $j=1$, (b) valence angles $j=2$, (c-e) different dihedral angles $j=3$, 4, and 5, respectively.
}
\label{fig02}
\end{figure}

The Hamiltonian of one molecule can be written as
\begin{equation}
H_0=\sum_{i=1}^{N_0}\Big[\frac12M_i(\dot{\bf u}_i,\dot{\bf u}_i)+P_i\Big],
\label{f1}
\end{equation}
where $i$ is the number of an atom, $N_0=6n^2$ is the number of atoms in the molecule, $M_i$ is the mass of the $i$th atom (there are $6n^2-6n$ inner carbon atoms of mass $12m_p$ and $6n$ edge carbon atoms of mass $13m_p$), ${\bf u}_i=(x_i(t),y_i(t),z_i(t))$ is the three-dimensional vector describing the position of $i$th atom at the time $t$. The term $P_i$ describes the interaction of the carbon atom with the index $i$ with the neighboring atoms. We emphasize that the inner and edge atoms differ only in their masses, and their interaction with each other is described by the same potential. The potential depends on variations in bond length, bond angles, and dihedral angles between the planes formed by three neighboring carbon atoms and it can be written in the form
\begin{equation}
P=\sum_{\Omega_1}U_1+\sum_{\Omega_2}U_2+\sum_{\Omega_3}U_3+\sum_{\Omega_4}U_4+\sum_{\Omega_5}U_5,
\label{f2}
\end{equation}
where $\Omega_j$, with $j=1$, 2, 3, 4, 5, are the sets of configurations describing different types of interactions between neighbors. Members of these sets are shown in Fig.~\ref{fig02}, and all their rotated and mirrored versions should be taken into account.

Potential $U_1({\bf u}_n,{\bf u}_m)$ describes the energy due to change in the length of a valence bond between atoms with the indexes $n$ and $m$, as shown in Fig.~\ref{fig02}(a).
The potential $U_2({\bf u}_n,{\bf u}_m,{\bf u}_k)$
describes the deformation energy of the angle between the valence bonds
${\bf u}_n{\bf u}_m$, and ${\bf u}_m{\bf u}_k$, see Fig.~\ref{fig02}(b). Potentials $U_j({\bf u}_n,{\bf u}_m,{\bf u}_k,{\bf u}_l)$,
$j=3$, 4, and 5, describe the deformation energy associated with a change in the angle between the
planes ${\bf u}_n{\bf u}_m{\bf u}_k$ and ${\bf u}_l{\bf u}_k{\bf u}_m$, as shown in Figs.~\ref{fig02}(c-e), respectively.

We use the potentials employed in the modeling of the dynamics of large polymer macromolecules
\cite{Noid19914148,Sumpter1994} for the valence bond coupling,
\begin{equation}
U_1({\bf u}_1,{\bf u}_2)\!=\!\epsilon_1
\{\exp[-\alpha_0(\rho-\rho_0)]-1\}^2,~\rho\!=\!|{\bf u}_2-{\bf u}_1|,
\label{f3}
\end{equation}
where $\epsilon_1$ is the energy of the valence bond and $\rho_0$ is the equilibrium length of the bond; the potential of the valence angle is
\begin{eqnarray}
U_2({\bf u}_1,{\bf u}_2,{\bf u}_3)=\epsilon_2(\cos\varphi-\cos\varphi_0)^2,~~
\label{f4}\\
\cos\varphi=({\bf u}_3-{\bf u}_2,{\bf u}_1-{\bf u}_2)/
(|{\bf u}_3-{\bf u}_2|\cdot |{\bf u}_2-{\bf u}_1|),~~
\nonumber
\end{eqnarray}
where the equilibrium value of the angle is $\cos\varphi_0=\cos(2\pi/3)=-1/2$;
the potential of the dihedral angle is
\begin{eqnarray}
\label{f5}
U_j({\bf u}_1,{\bf u}_2,{\bf u}_3,{\bf u}_4)=\epsilon_j(1+z_j\cos\phi),\\
\cos\phi=({\bf v}_1,{\bf v}_2)/(|{\bf v}_1|\cdot |{\bf v}_2|),\nonumber \\
{\bf v}_1=({\bf u}_2-{\bf u}_1)\times ({\bf u}_3-{\bf u}_2), \nonumber \\
{\bf v}_2=({\bf u}_3-{\bf u}_2)\times ({\bf u}_3-{\bf u}_4), \nonumber
\end{eqnarray}
where the sign $z_j=1$ for $j=3,4$ (the equilibrium value of the torsional angle $\phi$ is $\phi_0=\pi$) and $z_j=-1$ for $j=5$ ($\phi_0=0$).

The values of the potential parameters are $\epsilon_1=4.9632$~eV, $\rho_0=1.418$~{\AA}, $\alpha_0=1.7889$~\AA$^{-1}$, $\epsilon_2=1.3143$~eV, and $\epsilon_3=0.499$~eV. They are found from the frequency spectrum of small-amplitude oscillations of a graphene sheet~\cite{Savin08}. According to previous study~\cite{Gunlycke08}, the energy $\epsilon_4$ is close to
the energy $\epsilon_3$, whereas  $\epsilon_5\ll \epsilon_4$ ($|\epsilon_5/\epsilon_4|<1/20$). Therefore, we set $\epsilon_4=\epsilon_3=0.499$~eV and assume $\epsilon_5=0$, the latter means that we
omit the last term in the sum Eq.~(\ref{f2}).
More detailed discussion and motivation of our choice of the interaction potentials
Eqs.~(\ref{f3}-\ref{f5}) can be found in earlier publication~\cite{Savin10}.

The interaction of two coronene molecules is described by the potential
\begin{equation}
W({\bf X}_1,{\bf X}_2)=\sum_{i=1}^{N_0}\sum_{j=1}^{N_0}V(r_{ij}),
\label{f6}
\end{equation}
where the $3N_0$-dimensional vector ${\bf X}_k=\{ {\bf u}_{k,i}\}_{i=1}^{N_0}$ $(k=1,2)$
defines the coordinates of atoms of the $k$-th molecules (vector ${\bf u}_{k,i}$ specifies
the coordinates of the $i$-th atom of the $k$-th molecule), $r_{ij}=|{\bf u}_{2,j}-{\bf u}_{1,i}|$ is the distance between atoms. Nonvalence interactions of the carbon atoms are described by the (6,12) Lennard-Jones potential
\begin{equation}
V(r)=\epsilon_c\{[(r_c/r)^6-1]^2-1\},
\label{f7}
\end{equation}
where $\epsilon_c=0.002757$~eV, $r_c=3.807$~\AA~ \cite{Setton96}.

Hamiltonian of a chain of $N$ molecules (see Fig.~\ref{fig01}) can be presented in the form
\begin{eqnarray}
H=\sum_{n=1}^N\Big[\frac12({\bf M}\dot{\bf X}_n,\dot{\bf X}_n)+P({\bf X}_n)\Big]\nonumber\\
+\sum_{n=1}^{N-1}W({\bf X}_n,{\bf X}_{n+1})+\sum_{n=1}^{N-2}W({\bf X}_n,{\bf X}_{n+2}),
\label{f8}
\end{eqnarray}
where the first sum includes the kinetic and potential energies of $n$-th molecule. The second and the third sums describe the interaction between nearest and next-nearest molecules, respectively. Here the vector ${\bf X}_n=\{ {\bf u}_{n,i}\}_{i=1}^{N_0}$ specifies the coordinates of the atoms
of $n$-th molecule, ${\bf M}$ is the diagonal matrix of atom masses, $P({\bf X}_n)$ is the energy of $n$-th molecule, $W({\bf X}_n,{\bf X}_k)$ is the interaction energy of $n$-th and $k$-th molecules.

\section{The dispersion curves of small-amplitude oscillations}
\label{DispCurves}

Let us consider the structure of a symmetric (spiral) stack of planar $n$-coronene molecules with the symmetry axis parallel to the $z$ axis -- see Fig.~\ref{fig01}. In the ground state of such a chain, the atomic coordinates of each successive molecule are obtained from the coordinates of the previous molecule by translation along the $z$ axis by a shift $\Delta z$ and rotation around the same axis by an angle $\Delta\phi$. These are the shift and twist parameters:
\begin{eqnarray}
x_{n+1,j}&=&x_{n,j}\cos(\Delta\phi)+y_{n,j}\sin(\Delta\phi), \nonumber\\
y_{n+1,j}&=&-x_{n,j}\sin(\Delta\phi)+y_{n,j}\cos(\Delta\phi), \label{f9} \\
z_{n+1,j}&=&z_{n,j}+\Delta z,\nonumber\\
&&i=1,...,N_0,~n=0,\pm1,\pm2,... \nonumber
\end{eqnarray}

Thus, the energy of the ground state is a function of $3N_0$ coordinates of $N_0$ atoms of the first molecule ${\bf X}_1=\{ {\bf u}_{1,j}\}_{j=1}^{N_0}$, and the two geometry parameters, $\Delta z$ and $\Delta\phi$, where
${\bf u}_{1,j}=(x_{1,j},y_{1,j},z_{1,j})$ is the vector position of $j$th atom of the first molecule.

Finding the ground state reduces to the following minimization problem:
\begin{eqnarray}
E=P({\bf X}_1)+W({\bf X}_1,{\bf X}_2)+W({\bf X}_1,{\bf X}_3)\nonumber\\
\rightarrow
\min: \{ {\bf u}_{1,j}\}_{j=1}^{N_0},\Delta\phi,\Delta z.
\label{f10}
\end{eqnarray}
The problem (\ref{f10}) was solved numerically by the conjugate gradient method. The values of the shift $\Delta z$ and the twist $\Delta\phi$ steps of the chain of $n$-coronene molecules are presented in Table~\ref{tab1}.
\begin{table}[tb]
\caption{
Values of shift $\Delta z$ and twist $\Delta\phi$ parameters, maximum frequencies of out-of-plane $\omega_{op}$ and in-plane $\omega_{ip}$ vibrations, velocities of torsion $v_t$ and longitudinal $v_l$ sound for a spiral stack of $n$-coronene C$_{6n^2}$H$_{6n}$ molecules.
\label{tab1}
}
\begin{center}
\begin{tabular}{ccccccc}
 $n$     & $\Delta z$ (\AA) & $\Delta\phi$ ($^\circ$)  & $\omega_{op}$ (cm$^{-1}$) & $\omega_{ip}$ (cm$^{-1}$) & $v_t$ (m/s) & $v_l$ (m/s)\\
 \hline
 2       & 3.445            & 30.0  & 841.6 & 1549.3 & 217     & 3170\\
 3       & 3.411            & 18.6  & 883.7 & 1580.4 & 195     & 3449\\
 4       & 3.396            & 12.6  & 894.0 & 1591.3 & 250     & 3591\\
 \hline
\end{tabular}
\end{center}
\end{table}

A vertical chain of molecules is a multistable system. Numerical analysis shows that for $n$-coronene molecules with $n\le 4$, the spiral structure defined by Eq.~(\ref{f9}) is the most energy-favorable ground state.

For analysis of small-amplitude oscillations of spiral chain it is convenient to use local cylindrical coordinates ${\bf v}_{n,j}=(v_{n,j,1},v_{n,j,2},v_{n,j,3})$,
given by the following expressions:
\begin{eqnarray}
x_{n,j}&=&x_{n,j}^0+v_{n,j,1}\cos(\phi_{n,j})+v_{n,j,2}\sin(\phi_{n,j}), \nonumber\\
y_{n,j}&=&y_{n,j}^0-v_{n,j,1}\sin(\phi_{n,j})+v_{n,j,2}\cos(\phi_{n,j}), \label{f11}\\
z_{n,j}&=&z_{n,j}^0+v_{n,j,3}, \nonumber
\end{eqnarray}
with ${\bf u}_{n,j}^0=(x_{n,j}^0,y_{n,j}^0,z_{n,j}^0)$, ($n=0,\pm1,\pm2,...$; $j=1,...,N_0$)
being coordinates of the atoms in the helix ground state, and $\phi_{n,j}$ being angular coordinate
of the atom $(n,j)$.
With these new coordinates the Hamiltonian of the molecular chain Eq.~(\ref{f8}) has the following form
\begin{equation}
H=\sum_n\Big[\frac12({\bf M}\dot{\bf v}_n,\dot{\bf v}_n)+P({\bf v}_n,{\bf v}_{n+1},{\bf v}_{n+2})\Big],
\label{f12}
\end{equation}
where ${\bf v}_n=\{(v_{n,j,1},v_{n,j,2},v_{n,j,3})\}_{j=1}^{N_0}$ is a $3N_0$-dimensional vector, ${\bf M}$ is $3N_0$-dimensional diagonal mass matrix.
\begin{figure}[tb]
\begin{center}
\includegraphics[angle=0, width=1.\linewidth]{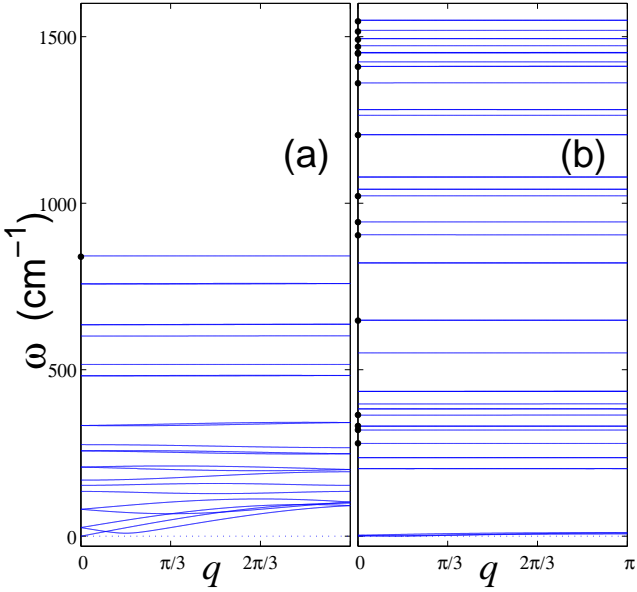}
\end{center}
\caption{\label{fig03}\protect
Structure of 72 dispersion curves of a spiral chain of coronene molecules C$_{24}$H$_{12}$ for (a)
out-of-plane and (b) in-plane vibrations.
Black dots denote modes leading to the formation of discrete breathers -- localized nonlinear oscillations of one molecule in the chain.
}
\end{figure}

From the Hamiltonian Eq.~(\ref{f12}) the following system of equations of motion can be derived:
\begin{eqnarray}
-{\bf M}\ddot{\bf v}_n=P_1({\bf v}_n,{\bf v}_{n+1},{\bf v}_{n+2})\nonumber\\
+P_2({\bf v}_{n-1},{\bf v}_{n},{\bf v}_{n+1})+P_3({\bf v}_{n-2},{\bf v}_{n-1},{\bf v}_{n}),
\label{f13}
\end{eqnarray}
where $P_i({\bf v}_1,{\bf v}_2,{\bf v}_3)=\partial P/\partial{\bf v}_i$, $i=1,2,3$.
Within the linear approximation, the system Eq.~(\ref{f13}) obtains the form
\begin{equation}
-{\bf M}\ddot{\bf v}_n=B_1{\bf v}_n+B_2{\bf v}_{n+1}+B_2^*{\bf v}_{n-1}+B_3{\bf v}_{n+2}+B_3^*{\bf v}_{n-2},
\label{f14}
\end{equation}
where the matrix elements are given as
$$
B_1=P_{11}+P_{22}+P_{33},~~
B_2=P_{12}+P_{23},~~
B_3=P_{13},
$$
and the partial derivative matrix is given as
$$
P_{ij}=\frac{\partial^2P}{\partial{\bf v}_i\partial{\bf v}_j}({\bf 0},{\bf 0},{\bf 0}),~~i,j=1,2,3.
$$
\begin{figure}[tb]
\begin{center}
\includegraphics[angle=0, width=1.\linewidth]{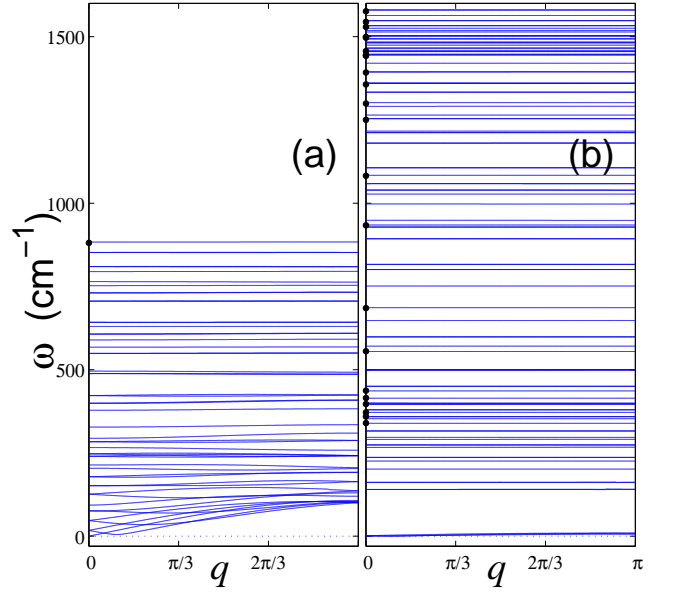}
\end{center}
\caption{\label{fig04}\protect
Structure of 162 dispersion curves for a spiral chain of circumcoronene molecules C$_{54}$H$_{18}$ for (a) out-of-plane and (b) in-plane vibrations. Black dots indicate modes that lead to the formation of discrete breathers -- localized nonlinear vibrations of one molecule in the chain.
}
\end{figure}

The solution to the system of linear equations Eq.~(\ref{f14}) can be found in the standard form
\begin{equation}
{\bf v}_n=A{\bf w}\exp[i(qn-\omega t)],
\label{f15}
\end{equation}
where $A$ is the linear mode amplitude, ${\bf w}$ is the eigenvector, $\omega$ is the phonon frequency  with the dimensionless wave number $q\in [0,\pi]$. Substituting Eq.~(\ref{f15}) into the system Eq.~(\ref{f14}), we arrive at the following $3N_0$-dimensional eigenvalue problem:
\begin{equation}
\omega^2{\bf Mw}={\bf C}(q){\bf w},
\label{f16}
\end{equation}
where Hermitian matrix
\begin{eqnarray}
{\bf C}(q)=B_1+B_2\exp(iq)+B_2^*\exp(-iq)\nonumber\\
                  +B_3\exp(2iq)+B_3^*\exp(-2iq)\nonumber.
\end{eqnarray}

Using the substitution ${\bf w}={\bf M}^{-1/2}{\bf e}$, problem Eq.~(\ref{f16}) can be rewritten in the form
\begin{equation}
\omega^2{\bf e}={\bf M}^{-1/2}{\bf C}(q){\bf M}^{-1/2}{\bf e}
\label{f17}
\end{equation}
where ${\bf e}$ is the normalized eigenvector, $({\bf e},{\bf e})=1$.

Thus, to obtain the dispersion curves $\omega_j(q)$, it is necessary to find the eigenvalues and eigenvectors of the Hermitian matrix Eq.~(\ref{f17}) of size $3N_0\times 3N_0$ for each fixed wavenumber $0\le q\le \pi$. As a result, we obtain $3N_0$ branches of the dispersion relation $\{\omega_j(q)\}_{j=1}^{3N_0}$.
\begin{figure}[tb]
\begin{center}
\includegraphics[angle=0, width=1.\linewidth]{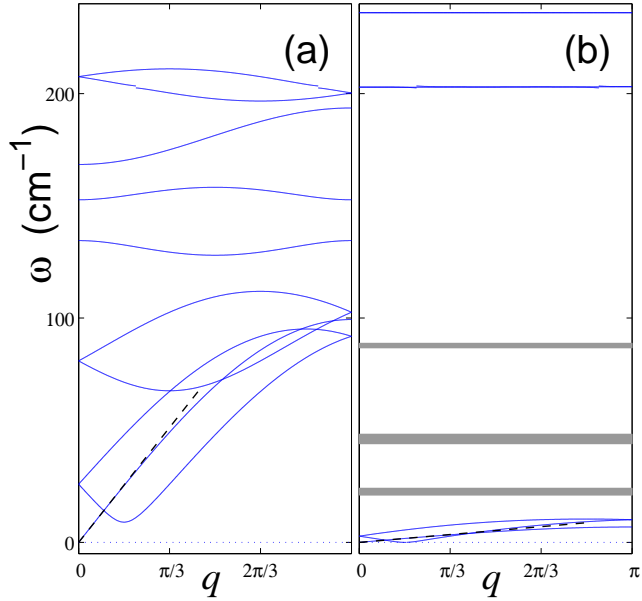}
\end{center}
\caption{\label{fig05}\protect
Dispersion curves in the low-frequency region for a spiral chain of coronene molecules C$_{24}$H$_{12}$ for (a) out-of-plane and (b) in-plane vibrations (three gray bands show the frequency spectrum of the rotobreathers). The dashed straight lines define the tangents to the dispersion curves emerging from the zero point, corresponding to the velocities of the longitudinal $v_l$ and torsion $v_t$ sound.
}
\end{figure}

The planar structure of molecules in a spiral chain leads to the division of its small-amplitude vibrations into two-classes: out-of-plane vibrations, when atoms vibrate orthogonally to the molecular plane (all atoms move along the $z$ axis) and in-plane vibrations (all atoms move in the $xy$ plane). Two thirds of the branches correspond to in-plane vibrations, while only one-third corresponds to out-of-plane vibrations.
The dispersion curves are shown in Figs.~\ref{fig03} to \ref{fig05}.

For the spiral chain of coronene molecules C$_{24}$H$_{12}$, the dispersion curves of out-of-plane vibrations, see Fig.~\ref{fig03}(a) and Fig.~\ref{fig05}(a), lie in the frequency range $0\le\omega\le\omega_{op}$, with the maximum frequency $\omega_{op}=842$~cm$^{-1}$. One dispersion curve $\omega_l(q)$ starts from the origin ($q=0$, $\omega=0$), it describes the displacement of planar molecules along the chain axis without internal deformations (longitudinal acoustic vibrations of the chain). The tangent of this dispersion curve at the origin gives the velocity of longitudinal sound waves
$$
v_l=\Delta z\lim_{q\rightarrow 0}\frac{\omega_l(q)}{q}.
$$

The dispersion curves of in-plane oscillations, see Fig.~\ref{fig03}(b) and Fig.~\ref{fig05}(b), lie in the frequency range $0\le\omega\le\omega_{ip}$ with the maximum frequency $\omega_{ip}=1549$~cm$^{-1}$. One dispersion curve $\omega_t(q)$ starts from the origin and describes torsional acoustic oscillations (rotation of planar molecules around the chain axis). The speed of long-wave torsional vibrations (speed of torsional sound) is
$$
v_t=\Delta z\lim_{q\rightarrow 0}\frac{\omega_t(q)}{q}.
$$

In addition, one dispersion curve approaches the $q$ axis tangentially. This curve describes the optical bending vibrations of the chain. The frequency spectrum of in-plane oscillations is characterized by the presence of a gap in the low-frequency region. For a chain of coronene molecules, the gap is from 10 to 203~cm$^{-1}$ [see Fig.~\ref{fig05}(b)], and for a chain of circumcoronene molecules, from 9 to 141~cm$^{-1}$ [see Fig.~\ref{fig04}(b)].

The values of the maximum frequencies $\omega_{op}$, $\omega_{ip}$ and the speeds of sound $v_l$, $v_t$ are given in Table~\ref{tab1}. As can be seen from the table, the speed of longitudinal sound is 15 times greater than the speed of torsional sound.
\begin{figure}[tb]
\begin{center}
\includegraphics[angle=0, width=1.\linewidth]{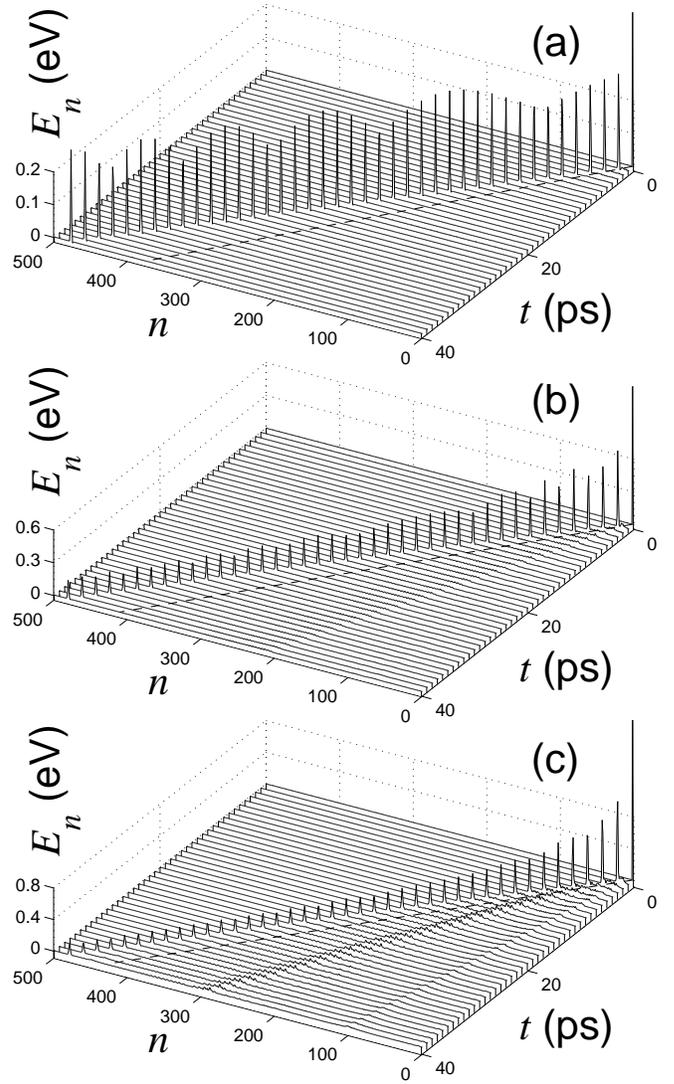}
\end{center}
\caption{\label{fig06}\protect
Formation of a supersonic acoustic soliton in a spiral chain of (a) coronene, (b) circumcoronene, and (c) dicircumcoronene molecules produced by longitudinal local compression at the end of the chain with amplitude $a_z=0.4$~{\AA}. The distribution of energy in the chain $E_n(t)$ at different times is shown. The number of molecules in the chain is $N=500$. The dotted lines show the trajectory of motion with the velocity of longitudinal sound $v_l$ to demonstrate the supersonic motion of solitons.
}
\end{figure}

\section{Acoustic solitons}
\label{AcousticSolitons}

The interaction of neighboring planar molecules is determined by the sum of interactions of all pairs of their atoms Eq.~(\ref{f6}), which are described by the Lennard-Jones potential Eq.~(\ref{f7}). The Lennard-Jones potential at small interatomic distances is characterized by the hard-type anharmonicity. Therefore, one can expect the possibility of propagation of compressive longitudinal acoustic solitons moving at a speed exceeding the velocity of longitudinal sound $v_l$.

To test the existence of supersonic acoustic solitons, we simulate the propagation of initial local longitudinal compression along a chain of molecules. Consider a spiral chain of $N=500$ molecules. Let us take the ground state of the chain and at $t=0$ shift the first two molecules along the $z$ axis by $a_z$. As a result, local longitudinal compression occurs at the end of the chain. Having fixed the position of these two molecules in the shifted state, let us consider the propagation of local compression along the chain.
\begin{figure}[tb]
\begin{center}
\includegraphics[angle=0, width=1.\linewidth]{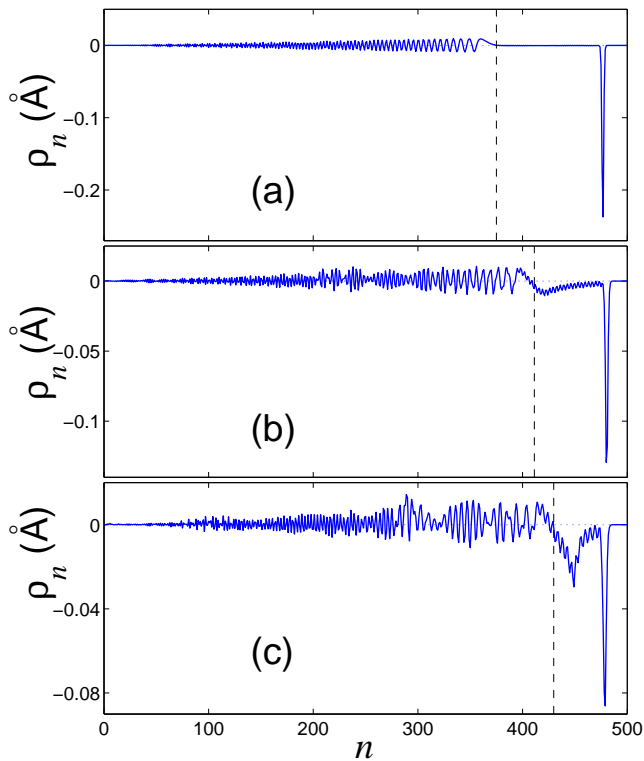}
\end{center}
\caption{\label{fig07}\protect
Distribution of longitudinal compression during the motion of an acoustic soliton along a chain of $N=500$ molecules of (a) coronene, (b) circumcoronene, (c) dicircumcoronene. The distribution of relative longitudinal displacements $\rho_n$ of chain molecules at time $t=40$~ps is shown for the amplitude of the initial local compression of the chain end $a_z=0.4$~\AA. The vertical dotted lines show the position of the front of the acoustic phonon wave packet propagating with the velocity $v_l$.
}
\end{figure}
\begin{figure}[tb]
\begin{center}
\includegraphics[angle=0, width=1.\linewidth]{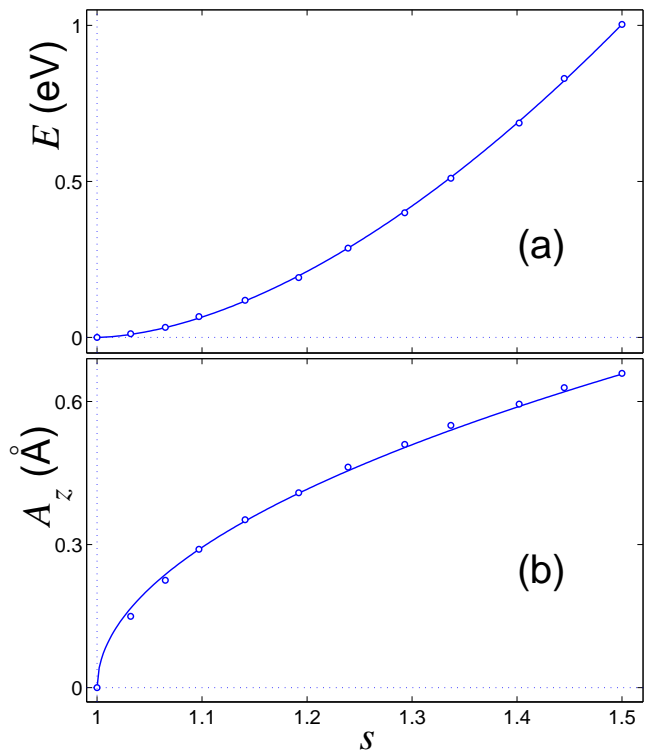}
\end{center}
\caption{\label{fig08}\protect
Dependence of (a) energy $E$ of an acoustic soliton and (b) longitudinal compression of the chain $A_z$ produced by an acoustic soliton propagating in a chain of coronene molecules on its dimensionless velocity $s=v/v_l$. Markers show numerical values, solid curves show approximations obtained by the least squares method $E(s)=3.36(s-1)^{1.7}$~eV and $A_z(s)=0.93(s -1)^{0,5}$~\AA.
}
\end{figure}

To simulate the dynamics of a chain with fixed ends, we numerically integrate the system of equations of motion corresponding to the Hamiltonian of the chain Eq.~(\ref{f8})
\begin{eqnarray}
{\bf M}\ddot{\bf X}_n&=&-\frac{\partial H}{\partial {\bf X}_n},~
n=3,4,...,N-2, \label{f18}\\
\dot{\bf X}_n&\equiv& {\bf 0},~~n=1,2,N-1,N, \nonumber
\end{eqnarray}
with the initial conditions
\begin{eqnarray}
{\bf X}_n(0)&=&{\bf X}_n^0+a_z{\bf e}_z,~~n=1,2 \nonumber\\
{\bf X}_n(0)&=&{\bf X}_n^0,~~n=3,4,...,N, \label{f19}\\
\dot{\bf X}_n(0)&=&{\bf 0},~~n=1,2,....,N,\nonumber
\end{eqnarray}
where the $3N_0$-dimensional vector ${\bf X}_n=\{(x_{n,j},y_{n,j},z_{n,j})\}_{j=1}^{N_0}$ defines the coordinates of the atoms of $n$-th molecule, vectors $\{ {\bf X}_n^0\}_{n=1}^N$ defines ground state of molecular chain, ${\bf e}_z$ is a unit vector directed along the $z$ axis, $a_z>0$ is the amplitude of the initial compression of the chain end.

Numerical integration of the system of equations of motion (\ref{f18}) showed that the initial longitudinal compression of the chain edge with an amplitude $a_z\le 0.6$~\AA~ for coronene molecules always leads to the formation of a supersonic acoustic soliton and a subsonic wave packet of long-wavelength longitudinal acoustic phonons -- see Fig.~\ref{fig06} (a) and \ref{fig07} (a). A local area of compression is formed in the chain, which moves along it with a constant supersonic speed $v>v_l$, keeping its shape. When moving, the soliton breaks away from the wave packet of phonons. This allows us to find its energy $E$ and the longitudinal compression of the chain $A_z$:
$$
E=\sum_nE_n,~A_z=\sum_n\rho_n,~\rho_n=\frac{1}{N_0}\sum_{j=1}^{N_0}(z_{n+1,j}-z_{n,j}-\Delta_z),
$$
where the summation is carried out only over the soliton localization region.

Dependencies of the soliton energy $E$ and chain compression $A_z$ produced by the soliton on its dimensionless velocity $s=v/v_l$ are shown in Fig.~\ref{fig08}. As can be seen from the figure, with increasing velocity, the soliton energy increases as $(s-1)^{1.7}$, and the compression as $(s-1)^{1/2}$.

In chains of circumcoronene and dicircumcoronene molecules, local longitudinal compression of the chain end also leads to the formation of a supersonic localized compression region. But the motion of this region is accompanied by the emission of phonons. As a result, the energy and velocity of the soliton decrease monotonically, see Figs.~\ref{fig06}(b,c) and \ref{fig07}(b,c). The larger the molecule, the more noticeable the emission of phonons. Therefore, it can be concluded that a chain of $n$-coronene molecules admits the existence of an exact acoustic soliton of longitudinal compression only for $n=2$, while for $n>2$ there is only a soliton-like excitation with a finite lifetime.
\begin{figure}[tb]
\begin{center}
\includegraphics[angle=0, width=1.\linewidth]{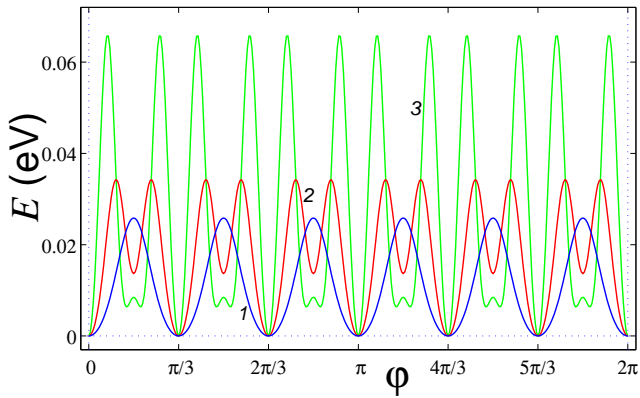}
\end{center}
\caption{\label{fig09}\protect
Change in the energy of the chain $E$ as the function of the rotation angle $\varphi$ of one molecule rotating around the $z$ axis in the chain of coronene, circumcoronene, and dicircumcoronene (curves 1, 2, and 3, respectively). Only one molecule rotates quasi-statically while the rest of the molecules remain in their equilibrium positions.
}
\end{figure}

\section{Rotobreathers}
\label{Rotobreathers}

The structure of planar molecules allows their rotation in chains around the $z$ axis. The $n$-coronene molecule has the shape of a regular hexagon, a rotation of one molecule by 60$^\circ$ will transfer the chain to an equivalent state. If we fix the positions of all molecules and rotate only one molecule as a rigid body, then the rotation potential $E(\varphi)$ (dependence of the chain energy on the angle of rotation of one molecule $\varphi$) can be obtained. This potential is a periodic function with period $\pi/3$, see Fig.~\ref{fig09}. In the approximation of absolutely rigid valence bonds, free rotation requires overcoming energy barriers of height 0.26, 0.34 and 0.66~eV for the chain of coronene, circumcoronene, and dicircumcoronene molecules, respectively. These barriers are overcome at molecular rotation frequencies above $\omega_0=2.19$, 1.11 and 0.87~cm$^{-1}$. Thus, the topology of the chain allows the existence of rotobreathers (localized rotations of molecules).

In the approximation of absolutely rigid molecules, their chains allow the existence of a rotobreathers with an infinite frequency spectrum
lying above frequency $\omega_0$. The $n$-coronene molecule is not an absolutely rigid body, it has $3N_0-6$ vibrational modes. The presence of internal vibrations in a rotator (in our case, a planar $n$-coronene molecule) leads to the appearance of band gaps (lacunae) in the frequency spectrum of the rotobreather~\cite{Savin22}. At frequencies within these band gaps, the rotation leads to resonance with the natural oscillations of the rotators and the emission of phonons. Therefore, the presence of internal vibrational modes in molecules should lead to a significant narrowing of the frequency spectrum of the rotobreather.
\begin{figure}[tb]
\begin{center}
\includegraphics[angle=0, width=1.\linewidth]{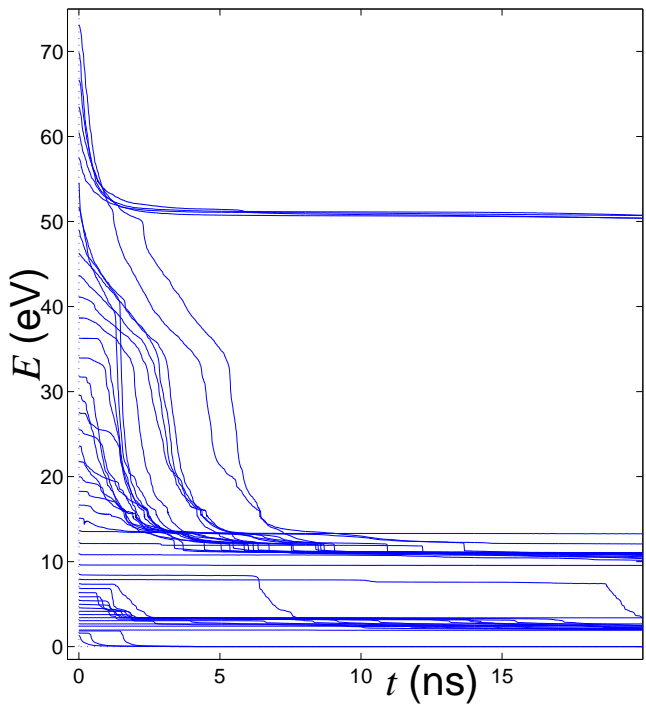}
\end{center}
\caption{\label{fig10}\protect
Change in time of the energy of one rotator in the chain of coronene molecules for different values of the initial rotation frequency of central molecule, varying in the range from $\omega=3$ to 22~cm$^{-1}$ with a step of 0.25~cm$^{-1}$.
}
\end{figure}
\begin{figure}[tb]
\begin{center}
\includegraphics[angle=0, width=1.\linewidth]{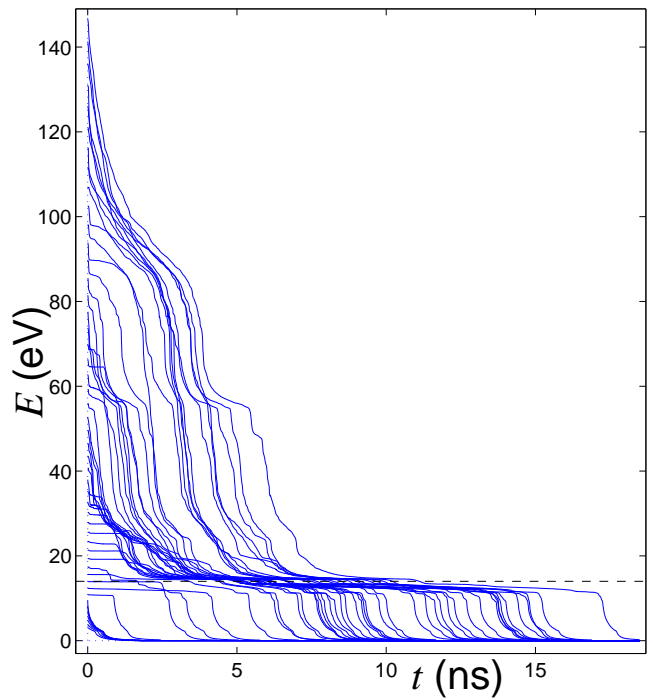}
\end{center}
\caption{\label{fig11}\protect
Change in time of the energy of one rotator in the chain of circumcoronene molecules for different values of the initial rotation frequency of central molecule, varying in the range from $\omega=2$ to 13.75~cm$^{-1}$ with a step of 0.25~cm$^{-1}$. The dashed line shows the energy corresponding to the rotation frequency $\omega=22.6$~cm$^{-1}$, at which the weakest phonon emission occurs.
}
\end{figure}

To find the rotobreather, we simulate the rotation of one molecule at different initial frequencies in a chain of $N=100$ molecules. A viscous friction at the ends of the chain is introduced, which ensures the absorption of phonons emitted by the rotator. To do this, we numerically integrate the system of equations of motion
\begin{eqnarray}
{\bf M}\ddot{\bf X}_n=-\frac{\partial H}{\partial {\bf X}_n},~~n=N_t+1,...,N-N_t,~~~~~\label{f20}\\
{\bf M}\ddot{\bf X}_n=-\frac{\partial H}{\partial {\bf X}_n}-\gamma{\bf M}\dot{\bf X}_n,
~~n\le N_t,~n>N-N_t \nonumber
\end{eqnarray}
with the friction coefficient $\gamma=1/t_r$, $t_r=10$~ps, $N_t=30$.

Let us take the ground state of the chain and excite the rotation of the central molecule $n_c=N/2$ with the frequency $\omega$, i.e. take the initial conditions in the form
\begin{eqnarray}
\{{\bf X}_n(0)={\bf X}_n^0\}_{n=1}^N,~\dot{\bf X}_n(0)={\bf 0},~~n\neq n_c ~~~~~~\label{f21}\\
\{\dot{x}_{n_c,j}=-\omega y_{n_c,j}^0,~\dot{y}_{n_c,j}=\omega x_{n_c,j}^0,~\dot{z}_{n_c,j}=0\}_{j=1}^{N_0}.
\nonumber
\end{eqnarray}
Thus, we set the rotation of one rotator in the chain with the initial energy
$$
E=\frac12\omega^2\sum_{j=1}^{N_0}M_j({x_{n_c,j}^0}^2+{y_{n_c,j}^0}^2).
$$

Friction at the ends of the chain will ensure the absorption of phonons emitted by the rotator. Therefore, depending on the value of the frequency $\omega$, the rotator either stops, having lost all the energy for phonon emission, or reaches a stationary rotation mode with a constant frequency without phonon emission (rotobreather mode). The change in the rotator energy $E$ for various initial values of the frequency $\omega$ in the chain of coronene and circumcoronene molecules is shown in Figs.~\ref{fig10} and \ref{fig11}, respectively.

As can be seen from Fig.~\ref{fig10}, for a chain of coronene molecules, there are only three frequency ranges at which a rotation at constant frequency of one molecule can occur without emitting phonons: [3.96, 4.54], [8.28, 9.09], and [16.33, 16.71]~cm$^{-1 }$. Thus, in the chain of coronene molecules, the rotobreather has a frequency spectrum consisting of only three narrow intervals, see also Fig.~\ref{fig05}(b), where the frequency spectrum of the rotobreather is shown by gray bands. Rotation with other frequencies leads to the emission of phonons.

Simulation of the dynamics of a rotator in a chain of circumcoronene molecules showed that rotobreathers do not exist in this chain. Here, at all values of the rotation frequency, the rotator emits phonons and completely loses energy, see Fig.~\ref{fig11}. There is only one frequency $\omega=22.6$~cm$^{-1}$ at which the radiation becomes less intense, but does not completely disappear. In a chain of dicircumcoronene molecules, the rotation of the rotator at all frequencies leads to an even stronger emission of phonons and no rotobreather is formed. The absence of a rotobreather in the chains of circumcoronene and dicircumcoronene molecules is explained by a denser frequency spectrum of natural vibrations of molecules. Here, in contrast to the coronene molecules, the rotation of the rotator at all frequencies resonates with the natural vibrations of the molecules.
\begin{figure}[tb]
\begin{center}
\includegraphics[angle=0, width=1.\linewidth]{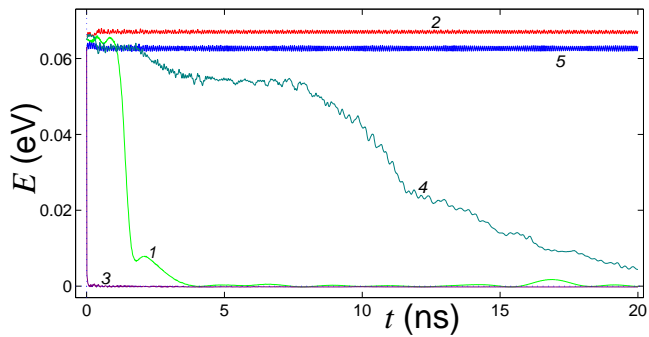}
\end{center}
\caption{\label{fig12}\protect
Dependence of the energy of vibrations of the central molecule of a chain of coronene molecules on time at the initial excitation of the $j$-th natural vibration:
(curve 1) $j=17$, $\omega_j=236.3$;
(curve 2) $j=21$, $\omega_j=278.8$;
(curve 3) $j=23$, $\omega_j=329.5$;
(curve 4) $j=33$, $\omega_j=435.2$;
(curve 5) $j=47$, $\omega_j=839.2$~cm$^{-1}$.
The initial atomic velocity used to excite the vibrational mode in the central molecule is $A=10$~\AA/ps.
}
\end{figure}

\section{Discrete breathers}
\label{DBs}

An isolated $n$-coronene molecule consists of $N_0=6n^2$ atoms. It has $3N_0-6$ natural oscillations with non-zero frequencies, $\{\omega_j\}_{j=7}^{3N_0}$. The first six eigenmodes have a zero frequency $\omega_1=...=\omega_6=0$, they correspond to the motion of a molecule as a rigid body (three translational and three rotational degrees of freedom). Eigenmodes with non-zero frequencies are of two types: $N_0-2$ out-of-plane vibrations, when atoms move orthogonally to the molecular plane, and $2N_0-4$ in-plane vibrations, when atoms move in the molecular plane.

The coronene molecule has 22 out-of-plane vibrations with frequencies 64.6, 117.7,..., 839.2~cm$ ^{-1}$ and 44 in-plane vibrations with frequencies 203.1, 236.3,..., 1546.2~cm$^{-1}$. The circumcoronene molecule has 52 out-of-plane vibrations with frequencies 32.3, 60.9,..., 881.0~cm$^{-1}$ and 104 in-plane vibrations with frequencies 140.5, 162.0,..., 1576.3~cm$^{-1}$. Let us check whether the excitation of a high-amplitude natural oscillation of one molecule can lead to the appearance of a discrete breather in the chain -- a nonlinear oscillation localized on one molecule.

To find discrete breathers, we simulate high-amplitude natural vibrations of one central molecule in a chain of $N=100$ molecules. At the ends of the chain, viscous friction is introduced, which ensures the absorption of phonons emitted by vibrations of the central molecule. The system of equations of motion Eq.~(\ref{f20}) is integrated numerically with the initial conditions
\begin{equation}
{\bf X}_n(0)={\bf X}_n^0,~\dot{\bf X}_n(0)=A{\bf e}_j\delta_{n,n_c},~n=1,...,N,\label{f22}
\nonumber
\end{equation}
where $A$ defines the magnitude of the initial velocity of atoms of the central molecule, ${\bf e}_j$ is the unit eigenvector of the $j$th eigenmode of an isolated molecule ($j=7$,...,$3N_0$), $n_c=N/2$. The value of $A$ determines the vibrational energy of the molecule and it is chosen sufficiently large to enter the regime of anharmonicity.

The dependencies of the vibrational energy of the central molecule on time $t$ are shown in Fig.~\ref{fig12}. Numerical integration of the system of equations of motion Eq.~(\ref{f20}) with the initial conditions Eq.~(\ref{f22}) showed that three dynamics scenarios are possible: very fast damping of oscillations (see Fig.~\ref{fig12}, curve 3), slow damping (curves 1 and 4) and the formation of undamped oscillations (curves 2 and 5). The first two scenarios are typical for out-of-plane vibrations, the last one -- for in-plane vibrations.
\begin{figure}[tb]
\begin{center}
\includegraphics[angle=0, width=1.\linewidth]{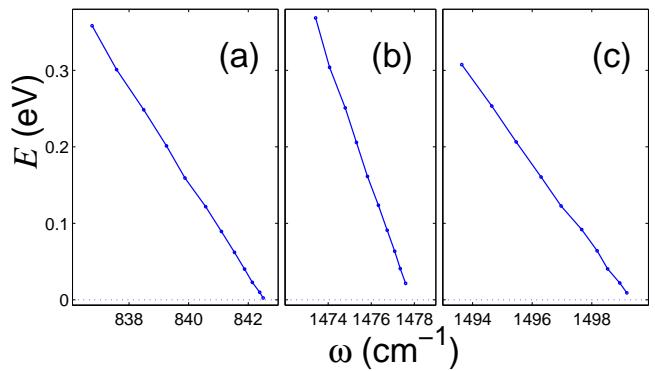}
\end{center}
\caption{\label{fig13}\protect
Dependence of the energy $E$ on the frequency $\omega$ for a discrete breather based on the $j$ eigenmode of the coronene molecule: (a) $j=47$, $\omega_j=839.2$; (b) $j=67$, $\omega_j=1470.0$ (c) $j=69$, $\omega_j=1491.3$~cm$^{-1}$.
}
\end{figure}

The frequencies of the resulting discrete breathers are shown by black dots in Figs.~\ref{fig03} and \ref{fig04}. Of all the out-of-plane eigenmodes, only the oscillation with the maximum frequency can lead to the formation of a discrete breather. For a chain of coronene molecules out of 44 in-plane vibrations 24 can lead to the formation of a discrete breather, and for a chain of circumcoronene molecules out of 104 in-plane vibrations 31 produce discrete breathers.

The undamped vibrations are localized strictly on one molecule. The oscillations are anharmonic, their frequency depends on the amplitude. A characteristic feature of localized oscillations (discrete breathers) is a linear decrease in their frequency with increasing energy, see Fig.~\ref{fig13}. As the oscillation amplitude increases, the energy of the breather increases and the frequency decreases. Thus, $n$-coronene chains support gap discrete breathers with a soft type of anharmonicity. The energy of a discrete breather in a chain of coronene molecules can reach 0.37~eV, and the width of the frequency spectrum can reach 6~cm$^{-1}$.

\section{Conclusions}
\label{Conclusions}

The linear phonon spectrum and nonlinear spatially localized excitations, such as acoustic solitons, rotobreathers, and discrete breathers in chains of $n$-coronene molecules, are studied by the method of molecular dynamics. Three members of the $n$-coronene were considered, namely coronene, circumcoronene and dicircumcoronene ($n=2$, 3 and 4 respectively). These molecules include respectively $N_0=24$, 54, and 96 carbon atoms and have $3N_0-6$ vibrational degrees of freedom.

The size of molecules plays an important role in chain dynamics. The spectra of low-amplitude vibrations of chains of coronene and circumcoronene molecules are shown in Figs~\ref{fig03} and \ref{fig04}, respectively. It can be seen that the maximum frequencies of out-of-plane and in-plane vibrations are approximately the same for chains of coronene and circumcoronene molecules, but the spectrum of the latter is denser, since the number of degrees of freedom is greater. The spectrum of a chain of dicircumcoronene molecules is even denser.

It was found that a chain of coronene molecules supports the propagation of acoustic compressive solitons, which practically do not emit energy when moving at supersonic speed, see Fig.~\ref{fig06}(a) and Fig.~\ref{fig07}(a). Similar excitations in chains of circumcoronene and dicircumcoronene molecules constantly lose energy, emitting low-amplitude phonons, see Fig.~\ref{fig06}(b,c) and Fig.~\ref{fig07}(b,c). This is because spiral chains have lower symmetry in the stacking of larger molecules and more channels to radiate energy due to the greater number of vibrational degrees of freedom.

A similar picture was observed for rotobreathers. Only in a chain of coronene molecules a single molecule can rotate with frequencies in certain ranges [shown in gray in Fig.~\ref{fig05}(b)], radiating no energy. In chains of circumcoronene and dicircumcoronene molecules, a molecule rotating at any frequency excites low-amplitude phonons, constantly loses its energy, and eventually stops rotating. The explanation lies in more resonances with a denser phonon spectrum in chains with larger molecules.

As for discrete breathers, they are supported by all three considered molecular chains. Discrete breathers are in the form of single molecule vibrating at large amplitude and radiating no energy. The frequencies of discrete breathers are marked with black dots in Figs.~\ref{fig03} and \ref{fig04} for chains of coronene and circumcoronene molecules, respectively. A discrete breather with out-of-plane oscillations, see panels (a), is created only by the highest-frequency out-of-plane mode. On the other hand, a number of in-plane vibrational modes create discrete breathers, see panels (b). The frequency of discrete breathers decreases with an increase in their energy, i.e. soft-type anharmonicity is realized, see Fig.~\ref{fig13}.

The results presented in this study illustrate the role of the internal degrees of freedom of particles in the nonlinear dynamics of molecular chains.

\begin{center}
{\bf ACKNOWLEDGMENTS}
\end{center}
Computational facilities were provided by the Interdepartmental Supercomputer Center of the Russian Academy of Sciences. The work of A.V.S. (statement of the problem, numerical simulations, and writing the manuscript) was supported by the Russian Science Foundation, Grant No. 21-12-00229. S.V.D. thanks the financial support provided by the Grants Council of the President of the Russian Federation grant NSh-4320.2022.1.2 (discussion of the results, writing the manuscript).

\bibliography{coronen}

\end{document}